# Graphene as a transparent conductive support for studying biological molecules by transmission electron microscopy

R. R. Nair[1], P. Blake[2], J. R. Blake[2], R. Zan[1,3], S. Anissimova[1], U. Bangert[3], A. P. Golovanov[4], S. V. Morozov[5], T. Latychevskaia[6], A. K. Geim[2], and K. S. Novoselov[1,*]

[1]School of Physics and Astronomy, the University of Manchester, Oxford Road, Manchester M13 9PL, United Kingdom
[2]Manchester Centre for Mesoscience and Nanotechnology, University of Manchester, Manchester M13 9PL, United Kingdom
[3]School of Materials, the University of Manchester, Manchester M1 7HS, United Kingdom
[4]Faculty of Life Sciences and Manchester Interdisciplinary Biocentre, The University of Manchester,131 Princess Street, Manchester M1 7DN, United Kingdom
[5]Institute for Microelectronics Technology, 142432 Chernogolovka, Russia
[6]Institute of Physics, University of Zurich, Winterthurerstrasse 190, CH-8057 Zurich, Switzerland
*Corresponding author: kostya@manchester.ac.uk

## Abstract

*We demonstrate the application of graphene as a support for imaging individual biological molecules in transmission electron microscope (TEM). A simple procedure to produce free-standing graphene membranes has been designed. Such membranes are extremely robust and can support practically any sub-micrometer object. Tobacco mosaic virus has been deposited on graphene samples and observed in a TEM. High contrast has been achieved even though no staining has been applied.*

## Introduction

Modern high-energy transmission electron microscopes (TEMs) operate with electrons of energy in range of about 80-200keV and are capable of delivering images at resolution of about 2Å, thus making the instruments suitable for direct visualization of molecules at atomic resolution. Traditionally, and since the very first experiments in TEM, the objects under study are placed on some thin, transparent for electron waves, amorphous carbon film of about 10nm thickness. Such carbon films however are not the most suitable support for imaging and resolving individual molecules or atoms, one of the goals of modern research. In the case of carbon films, the electron wave scattered by the supporting carbon film produces background noise, reducing the overall image contrast and obscuring the true image of the object. One solution to this problem is stretching the objects under study over holes in carbon film. However, this method is limited to certain shapes of objects and cannot be used for imaging small round-shape molecules. In an ideal world, the support for such unstained molecules would be a continuous structure-less electron-transparent film.

Graphene (*1-2*) has recently been suggested as a support film for samples in TEM. This material has a highly ordered and stable structure and is the strongest material currently known, which makes it mechanically an optimal support layer. It is also electrically conductive, ensuring an equipotential surface and eliminating the problem of the phase alteration of the passing electron wave; the latter is crucial importance for performing holography experiments (*3-4*). Finally, graphene is the ultimately thin material - it is just one layer of carbon atoms thick and thus, as thin as any traditional carbon support film could be. Its thickness is estimated to be about 0.34nm (*5*). Due to its high level of crystallinity, graphene has an ordered structure whose contribution to the image can be easily filtered out by blocking characteristic peaks in the Fourier domain (*6-7*). Overall, the properties of graphene discussed above make it an ideal support for samples in TEM.

Graphene has already been proven as a successful support film for imaging individual hydrogen atoms, carbon adsorbates (*6*) and gold nanoparticles (*7*) in TEM; $CoCl_2$ nanocrystals on a graphene sheet were measured in high resolution TEM mode (*8*). A hydrophilic modification of graphene, graphene oxide, was used to image 26S proteasome in TEM (*9-10*). In this letter, we show how unstained biological samples such as individual tobacco mosaic viruses (TMV) can be placed on graphene film and imaged in TEM.

## Sample preparation

Graphene membrane preparation procedure was as follows. Graphene flakes have been obtained from Graphene Industries (*11*). They were produced by mechanical exfoliation of high quality graphite on $Si/SiO_2$ substrate (300 nm of $SiO_2$) (*1, 12*) where graphene flakes could be easily identified (*13*), see Fig. 1. Next, 100 nm of poly-methyl methacrylate (PMMA) was spun on top. Removing the $SiO_2$ layer (by etching it away in KOH) released the PMMA layer (with graphene flakes attached to it) which was floating on the surface of KOH solution. After several rinsings with water, the PMMA layer was fished out on a Quantifoil grid (*14*) and successively dissolved in acetone. The TEM grid was dried in a critical point dryer, and transferred graphene flakes could be identified by Raman spectroscopy or visually by the presence of the impurities (Fig.2). Such graphene membranes produce very small contrast either for back-scattered electrons in SEM or for transmission electrons in TEM, at least for large enough (>1kV) acceleration voltage. We would like to mention that previously it has been proposed (*1*) to use a similar procedure to transfer graphene grown by chemical vapor deposition (CVD) on either Ni or Cu (*15-17*), which would allow one to cover the complete TEM grid with graphene (*18*).

Obtained graphene membranes are extremely robust and can be used to hold practically any object of submicrometer size. For our purpose we chose to deposit TMV for observation in TEM. TMV is the best known object for pioneering research in structural biology (*19-20*). It was the first virus to be directly visualized in TEM in 1939 (*21*), when its simple geometrical shape was identified; it is rodlike with length of about 300nm and diameter of about 18nm.

TMV has been obtained from LGC-ATCC (ATCC number PV-1, shipped freeze-dry) (*22*). Standard procedure was used to prepare suspension; (i) 0.1g of leaf tissue crashed together with 1ml of water; (ii) the suspension centrifuged for 1min at 14000rpm; and (iii) the supernatant solution filtered through 0.5μm filters. Graphene membranes were then dipped into the obtained suspension and dried under ambient conditions. No further processing (e.g., staining (*23*)) has been performed.

## Experimental results

For our TEM experiments we used a Technai F30 microscope (acceleration voltage 300kV). Graphene membranes have been found by occasional specks of dirt on its surface (Figs. 2 and 3) and identified as monolayers by the diffraction patterns observed under tilt (*24*). Closer inspection revealed the presence of TMV molecules on the surface [Figs. 3(a) and 3(b)]. Two types of contamination are evident on our graphene samples. One is patchy with a typical size of a few hundred nanometers and high-contrast features. This contamination is likely a result of the transfer procedure and, although it does not obscure our observation of nanometersized objects, could be avoided with a better cleaning procedure for PMMA. The second type of contamination shows a continuous gray background which provides a high contrast between graphene and an empty space [e.g., ripped hole in graphene, Fig. 3(a)]. Such strong contamination has not been observed on pristine graphene membranes (*24*) before depositing the virus (Fig. 2), so we attribute it to contamination from the organic residuals in the TMV suspension. Theoretically, better centrifuging and filtering should help in eliminating such contamination.

Despite the strong contamination, individual TMVs are clearly visible (even without staining) as rodlike objects and their length and diameter are estimated to be about 18nm in width and 300nm in length, both in perfect agreement with the description mentioned above. The contrast achieved in our experiments is of the order of 0.3. Such good visibility of the viruses can be explained by the fact that graphene itself provides extremely weak absorption, and the adsorbed contamination is still lower than that on the usual substrates (amorphous carbon film).

## Summary

In summary, we provided a simple recipe of preparation of graphene films with adsorbed biological molecules. As an example of our technique, we demonstrated images of individual unstained TMV on graphene support recorded in TEM.

## Acknowledgements

This work was supported by Engineering and Physical Sciences Research Council (U.K.), the Royal Society, the European Research Council (programs "Ideas," call Grant No. ERC-2007-StG and "New and Emerging Science and Technology," project "Structural Information of Biological Molecules at Atomic Resolution"), Office of Naval Research, and Air Force Office of Scientific Research. The authors are grateful to Nacional de Grafite for supplying high quality crystals of graphite. The authors are grateful to Hans-Werner Fink and Conrad Escher for valuable discussions.

# Figures

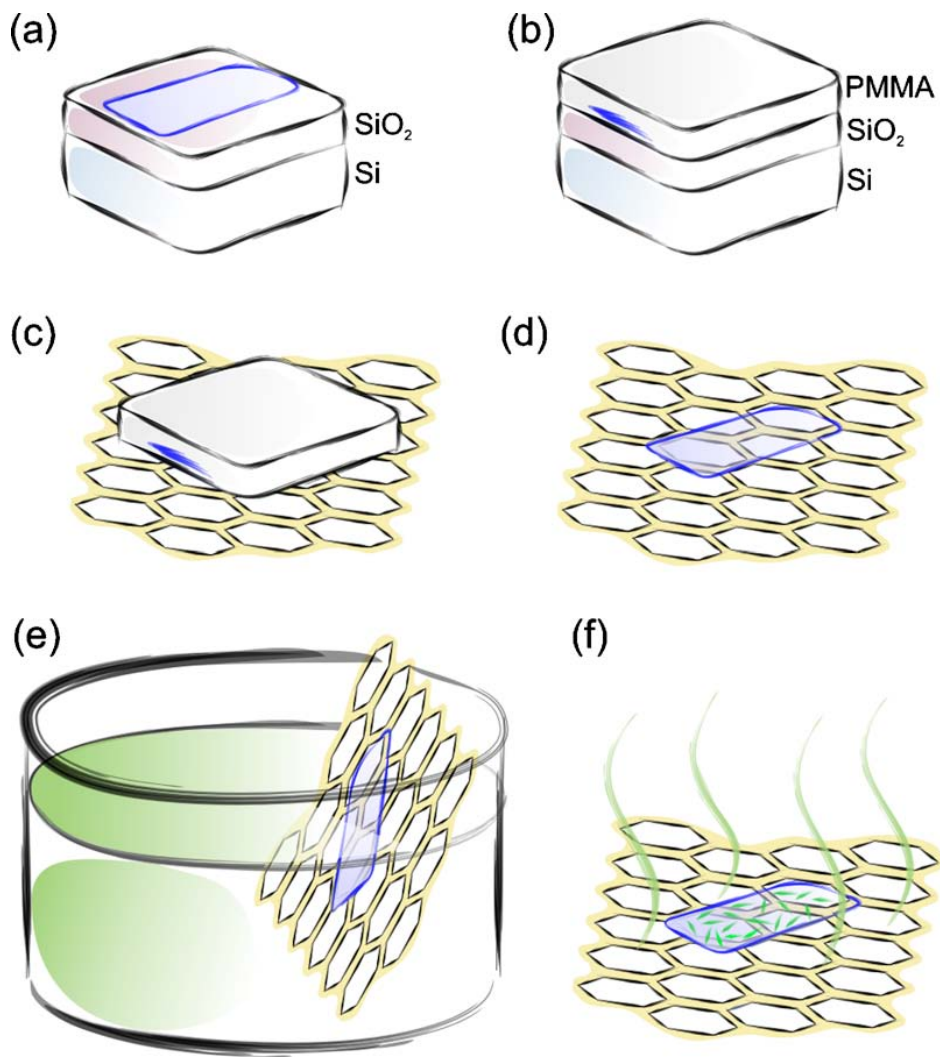

Fig. 1 Procedure for the preparation of graphene membranes [(a)-(d)] and deposition of biomolecules [(e)-(f)]: (a) graphene flakes are obtained on Si/SiO2 substrate by mechanical exfoliation method; (b) 100nm PMMA film is spun on top of graphene; (c) sacrificial $SiO_2$ layer is removed in KOH and the released PMMA layer with graphene flakes attached is fished out by Quantifoil TEM grid; (d) PMMA layer is removed by dissolving in acetone and the Quantifoil TEM grid is dried in a critical point dryer; (e) such membranes are then dipped into a solution containing biomolecules of interest; and (f) dried under ambient conditions, leaving biomolecules attached to graphene membranes.

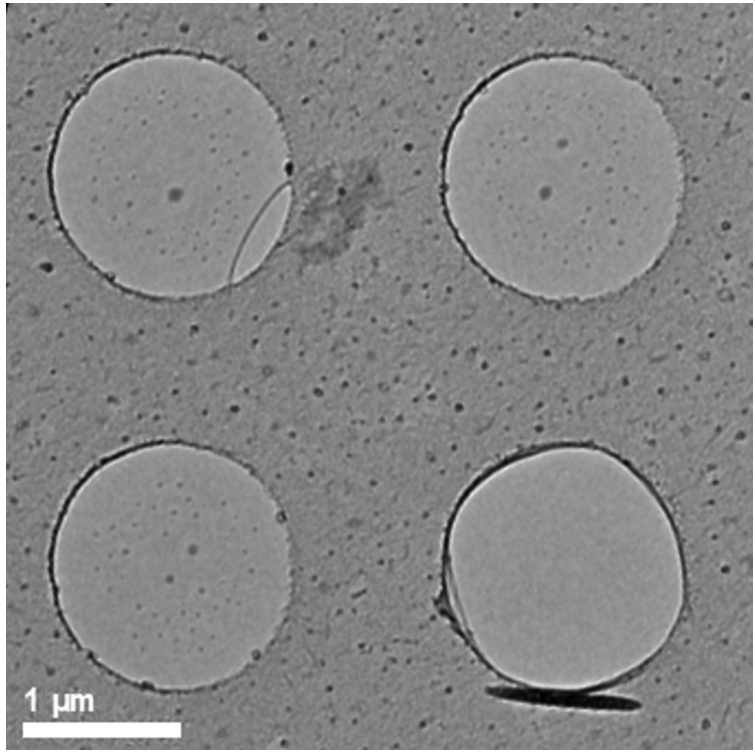

Fig. 2 TEM image of graphene on top of Quantifoil TEM grid. Graphene has been placed on amorphous carbon film with periodic array of holes (1.4μm in diameter). Here the three holes are covered with graphene, while one hole is left empty (bottom right).

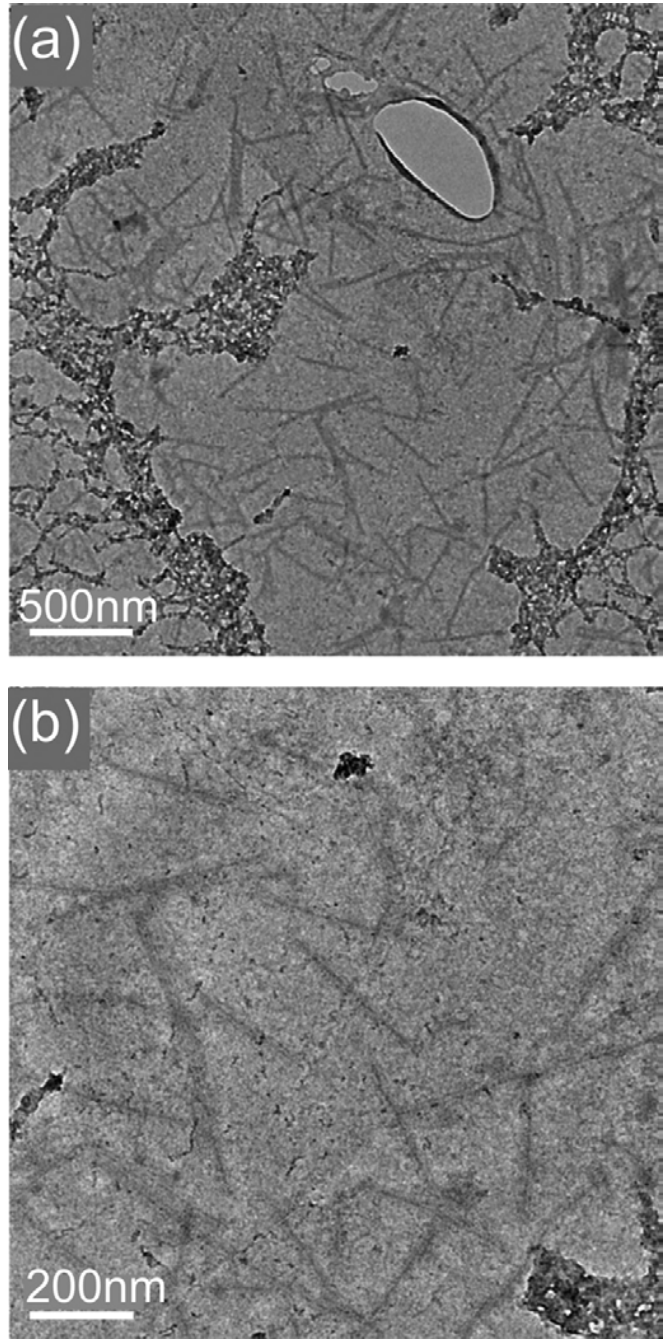

Fig. 3 TEM image of one of our graphene membranes with TMV on top. (a) and (b) are different magnifications.